\documentclass[ twocolumn, tighten]{aastex631}
\usepackage{graphicx}
\usepackage{natbib}
\usepackage{amsmath}
\usepackage{ mathrsfs }
\usepackage{subfigure}
\usepackage{epstopdf}
\usepackage{tipa}
\usepackage{comment}

\long\def\symbolfootnote[#1]#2{\begingroup%
\def\thefootnote{\fnsymbol{footnote}}\footnote[#1]{#2}\endgroup}

\newcommand{\beq}{\begin{equation}}
\newcommand{\eeq}{\end{equation}}
\newcommand{\bea}{\begin{eqnarray}}
\newcommand{\eea}{\end{eqnarray}}

\newcommand{\sigmac}{{\sigma_c}}

\newcommand{\Mdotc}{{\dot{M}_c}}
\newcommand{\Mc}{{M_c}}

\newcommand{\etaM}{{\eta_M}}

\newcommand{\epsin}{{\epsilon_{\rm in} }}

\newcommand{\xic}{{\xi_c}}

\newcommand{\SFRff}{{{\rm SFR}_{\rm ff}}}

\newcommand{\Mach}{{\cal M}}

\newcommand{\vchar}{v_{{\rm ch},w}}

\newcommand{\phiacc}{\varphi_{\rm acc}}

\newcommand{\Nfil}{N_{\rm fil}}

\newcommand{\mPlanck}{m_{\rm Pl}} 
\newcommand{\deltagr}{\delta_{\rm gr}}
\newcommand{\kappaCGS}{\kappa_{\rm cgs}}
\newcommand{\SigmaCGS}{\Sigma_{\rm cgs}}
\newcommand{\alphavir}{\alpha_{\rm vir}}
\newcommand{\xifil}{\zeta_{\rm fil}} 

\newcommand{\newtext}[1]{{#1}}

\begin{document}

\title{Intense star cluster formation: stellar masses, the mass function, and the fundamental mass scale} 
\shorttitle{IMF and the Fundamental Mass Scale}
\shortauthors{Matzner}

%\maketitle

\author{Christopher D. Matzner}
\affil{David A.\ Dunlap Dept.\ of Astronomy and Astrophysics, University of Toronto, 50 St. George Street, Toronto, Ontario, M5S 3H4, Canada}

\email{matzner@astro.utoronto.ca}

%\maketitle{}
\begin{abstract}
Within  the birth environment of a massive globular cluster, the combination of a luminous young stellar population and a high column density induces a state in which the thermal optical depth and radiation pressure are both appreciable.  In this state, the sonic mass scale, which influences the peak of the stellar mass function, is tied to a fundamental scale composed of the Planck mass and the mass per particle.  Thermal feedback also affects  the opacity-limited minimum mass and affects how protostellar outflows and binary fragmentation modify stellar masses. Considering the regions that collapse to form massive stars, we argue that thermal stabilization is likely to flatten the high-mass slope of the initial mass function.  Among regions that are optically thick to thermal radiation, we expect the stellar population to become increasingly top-heavy at higher column densities, although this effect can be offset by lowering the metallicity.   A toy model is presented that demonstrates these effects, and in which radiation pressure leads to gas dispersal before all of the mass is converted into stars. 
\end{abstract}

\section{Introduction}\label{S:Intro}

We wish to examine how environment affects the stellar initial mass function (IMF) during intense bursts of star cluster formation like those that give rise to massive globular clusters.  Along the way, we will explore link between the peak of the IMF and the fundamental mass scales of stellar evolution.  

 The IMF $\psi(m_*) \propto d\dot N_*/d m_*$ is remarkably consistent in many well studied environments \citep{2003PASP..115..763C}, featuring  a roughly log-normal distribution at low masses with a peak in $d\dot N_*/d \log m_*\propto m_* \psi(m_*)$ around $0.2M_\odot$ and a power law decline to high masses, $\psi \propto m_*^{-\alpha_{\rm high}}$ (with $\alpha_{\rm high} \simeq 2.35$ for $m_*>M_\odot$) which meets a cutoff somewhat above $10^2M_\odot$.    However, in some cases massive stars appear to be over-represented in the birth distribution. 
Examples from the inner Milky Way include the Arches \citep{2019ApJ...870...44H} and the young stars at the Galactic Center \citep{2005MNRAS.364L..23N,2013ApJ...764..155L}.   

\newtext{There are now many extragalactic examples as well.} An overabundance of massive stars is one plausible explanation for the unexpectedly numerous high-redshift ultraviolet sources discovered by JWST \citep{2023arXiv230811609F}.   Very massive stars ($m_*>100M_\odot$) are copious wind sources of matter processed by H fusion at $\sim 7.5\times10^7\,$K, which is enriched in elements, like N and Na, seen to pollute
N-enriched nebulae at moderate redshift ($z=2.4$: \citealt{2023ApJ...957...77P} and \citealt{2024arXiv240410755P}) and high redshift ($z=9.4$: \citealt{2024arXiv240608408S}; $z=8.3$: \citealt{2024arXiv240714201N}; $z=8.7$ and others: \citealt{2024A&A...681A..30M}).  \newtext{\cite{2024arXiv240719009T} infer that N-enriched nebulae are common in a large sample at $z>4$.} 
Indeed, there is evidence that the distinct  features of very massive stars, such as strong He\,II emission, are prominent in the spectra of some young starburst clusters \citep{2023A&A...679L...9V, 2023ApJ...958..194S}.    \cite{2024A&A...686A.185U} present evidence that very massive stars are over-represented in 13 systems between $z=2.2$ and $z=3.6$.   Similarly,  \citet{2023arXiv231102051C} determine that stars with $m_*>50M_\odot$ must be strongly overabundant to explain two-photon emission in nebular-dominated galaxies  at  $z=5.9$ and $z=7.9$.   
\newtext{Some of these may be examples of globular cluster formation, considering that the most massive globulars harbor secondary stellar populations enriched in N and other hot burning products \citep{2012A&ARv..20...50G}.  }

These phenomena bespeak the formation of a top-heavy IMF, either because the high-mass slope flattens (e.g., $\alpha_{\rm high} <2$) or because the peak mass or upper mass limit shifts upward, or some combination of these.  At the same time, the IMF may become bottom-light: \cite{2011AJ....142....8S} and \cite{2023MNRAS.521.3991B} infer that Galactic globular clusters formed a dearth of relatively low-mass stars.

Because massive globular clusters are expected to form in one or more intense bursts, under extreme conditions in terms of the column density $\Sigma$, velocity dispersion $\sigma$, and other parameters, it is logical to ask how these conditions might lead to the over-production of massive stars.  \cite{2023ApJ...959...88D} argues that this could be the consequence of collisions between molecular cores in the high-$\Sigma$ environment.   \cite{2024ApJ...966...48C}, considering the Central Molecular Zone,  attribute the flatter IMF to an impulsive injection of turbulence within the theoretical framework of \citet{2013ApJ...770..150H}.   We, in contrast, will examine the thermal evolution of these extreme events and ask what implications it has for the stellar mass distribution.  

We will focus on conditions representative of massive globular cluster formation, involving systems with total mass $M = 10^6 M_6 M_\odot$ and column density $\Sigma = \SigmaCGS\,{\rm g\,cm^{-2}}$ where $M_6$ and $\SigmaCGS$ are both of order unity or greater.  We note that the Milky Way globular clusters compiled in the 2010 edition of the \citet{1996AJ....112.1487H} catalog have half-light radii of several pc nearly independently of mass, indicative of $\SigmaCGS\sim 3 M_6$ within that radius.   Because of the internal and tidal dynamical interactions that affect cluster evolution \citep[e.g.,][]{2010MNRAS.406.2000M} this may not faithfully represent the birth conditions, which presumably had higher $M$ and $\Sigma$.  
An upper bound comes from  \cite{hopkins2010maximum}, who find   $\SigmaCGS\leq20$ for all known stellar systems. 

Considering only large values of $M$ and $\Sigma$ simplifies our analysis in two ways.   First, as the  mean hydrogen density is $10^{4.1} \SigmaCGS^{3/2} M_6^{-1/2}$\,cm$^{-3}$ and the H density of a sonic core  (see \S\,\ref{S:Msonic}) is $\sim 10^{6.8}\SigmaCGS^2$\,cm$^{-3}$,
gas and dust are likely to be strongly thermally coupled \citep{2001ApJ...557..736G}.  \newtext{(We examine the conditions for thermal coupling in \S\,\ref{SS:validity}.)}  So long as they are, we can use the dust temperature estimated from radiative transfer as a proxy for the gas temperature, rather than addressing the details of heating and cooling that are critical for analyses of IMF formation in more typical environments \citep[e.g.,][]{2008ApJ...681..365E}.

Second, the high columns and velocity dispersions ($\sigma = 10.3 \alphavir^{1/2} M_6^{1/4} \SigmaCGS^{1/4}$\,km\,s$^{-1}$, where $\alphavir = 5 \sigma^2 R/GM$ is the \citealt{1992ApJ...395..140B} virial parameter)  imply that most channels of  stellar feedback  -- protostellar outflows, photo-ionized gas pressure, shocked stellar wind pressure,  direct starlight momentum, and supernovae -- will fail to affect motions on scale of the region (see \citealt{2010ApJ...710L.142F}, \citealt{2015ApJ...815...68M}, and \citealt{2024ApJ...967L..28M}, for instance).  For this reason we will only consider protostellar outflows  insofar as they affect the sub-regions that collapse into individual stellar systems and subgroups.  On the cloud scale, the only relevant mode of dynamical feedback involves radiation pressure gradient, which in \S\,\ref{S:ingredients} we encapsulate in terms of the Eddington ratio. 

\subsection{The fundamental mass scale and stellar evolution} 
A central touchstone for our study of the origin of stellar masses is the {\em fundamental mass scale} \[ \mPlanck^3/\hat \mu^2\] in which $\mPlanck = (c\hbar/G)^{1/2}$ is the Planck mass and $\hat\mu$ is the relevant mass per particle.  In practice  $\hat\mu$ is within a factor of a few of the proton mass $m_p$,  and so the fundamental scale is within an order of magnitude of $\mPlanck^3/m_p^2 = 1.85M_\odot$.   

All of the critical mass scales in stellar evolution are cousins of the fundamental scale. These 
include  the mass limits for steady [D or H] burning, 
\[ [m_{*D}, m_{*H}] \simeq 2.8 \left(\frac{\mu_{*e}}{m_e}\frac{\Psi_{*[D,H]}}{c^2} \right)^{3/4}\, \frac{\mPlanck^3}{\mu_{*e}^2} \] \[ ~~= [0.02, 0.08]\, M_\odot, \]  
\citeauthor{1931ApJ....74...81C}'s (\citeyear{1931ApJ....74...81C}) limiting mass of a white dwarf, 
\[m_{\rm Ch} = 3.10\, \mPlanck^3/\mu_e^2 \simeq 1.43\,  M_\odot, \]
and the stellar mass for which radiation and gas pressures are equal, which in \citeauthor{1920Obs....43..341E}'s (\citeyear{1920Obs....43..341E}) standard model is 
\[ m_{*\rm rad} = 27.5\, \mPlanck^3/\mu_*^2 \simeq 140 \,M_\odot. \] 
The evaluation assumes cosmic composition. 
Note that a star's diffusive luminosity approaches Eddington's limit $L_{*\rm Edd} = 4\pi G m_* c/\kappa_{\rm es}$ for stellar masses $m_*>m_{*\rm rad}$.   

(In the  expressions above, $\mu_* \simeq 4m_p/(3+5X)\simeq 0.6m_p$ is the mean molecular weight, 
and $\mu_{*e} \simeq 2m_p/(1+X) \simeq 1.17m_p$ is the mass per electron,
within a star  of hydrogen mass fraction $X\simeq0.7$;  $\mu_e\simeq 2 m_p$ is the mass per electron in a white dwarf,  $[\Psi_{*D}, \Psi_{*H}]\simeq GM/([5,0.8]R_\odot)$ is the  gravitational potential depth at the surface of a fully convective star that stably burns [D or H] in its core, and $\kappa_{\rm es} = 0.2(1+X)$cm$^2$\,g$^{-1}$ is the stellar electron scattering opacity.  Coulomb corrections are neglected.) 

The formulae for $m_{*D}$, $m_{*H}$, and $m_{\rm Ch}$ include $\hbar$ and $\mu_e$ because of the importance of electron degeneracy, and $c$ because of the normalization of $\Psi_{*,[D,H]}$ (in the burning limits) or because the electrons are relativistic (in Chandrasekhar's limit).  

It is worth commenting on how the fundamental mass arises in $m_{*\rm rad}$. Here the hydrostatic pressure ($\sim GM^2/R^4$) is balanced in equal parts by 
\newtext{radiation pressure ($\sim (kT)^4/(\hbar c)^3$) and gas pressure ($\sim MkT/\mu R^3$).  The coincidence of these pressures sets $M\sim \mPlanck^3/\mu^2$, with a numerical prefactor determined by the gas-radiation pressure ratio and the polytropic structure. }

A similar situation arises in star formation, when radiation dynamics combine with gas pressure to set a critical mass.  In the prototypical example, \cite{1976MNRAS.176..483R} has shown that the minimum Jeans mass in a fragmenting cloud
 is related to the fundamental mass scale of molecular gas, with a prefactor that accounts for the temperature of the environment and aspects of radiative transfer (see \S\,\ref{S:Mmin}).  
  We are especially interested in the possibility that other mass scales, such as the IMF peak (the maximum of $m_*\psi \propto d\dot N_*/d\log m_*$) and the upper mass cutoff, are also related to the fundamental mass.    
  \newtext{As we shall see, these questions are especially pertinent to globular cluster formation.}

\section{Ingredients} \label{S:ingredients} 
To begin, we gather a few useful ingredients: a simplified radiation transfer solution, an estimate of the relevant luminosities and opacities that enter the radiation transfer calculation, and a rule for extrapolating turbulent motions to small scales.  

To highlight the analytical limits for the radiative transfer problem, we adopt a simplified model for the dust temperature distribution $T(r)$ arising due to the reprocessing by dust of a stellar luminosity $L(r)$ within a region of radius $r$ that is optically thick to starlight:
\begin{equation} \label{eq:Tapprox}
T^4 = \frac{L(r)}{\pi r^2 a_r c } f(\tau),
\end{equation} 
where $a_r = \pi^2 k^4/15\hbar^3c^3$ is the radiation constant and 
\begin{equation} 
 f(\tau) \rightarrow \left\{ \begin{array}{lc} 1/\tau & \tau\ll 1 \\
%&\\ 
\tau/\phi_\tau & \tau\gg 1 \end{array} \right. 
\end{equation} 
 where 
 \begin{equation} 
 \tau(r) = \kappa(T)f_g  \Sigma(r)
 \end{equation} 
 is the thermal optical depth at $r$.  Here $\Sigma(r)=M(r)/\pi r^2$ is the mean column and $f_g = 1-f_*$ is the gas fraction.   An analysis 
based on the diffusion approximation within a spherical power law density profile $\rho\propto r^{-k_\rho}$ indicates 
 $\phi_\tau^{1/4}\simeq [(16/3)(5-k_\rho)/(3-k_\rho)]^{1/4} \simeq 1.63$ for representative density slopes $1<k_\rho<2$.  
  
For the sake of clarity, except in \S\,\ref{SS:toymodel} we shall define the thick and thin limits of each expression of interest, rather than introducing in interpolating function 
or some more accurate treatment.     Equation (\ref{eq:Tapprox}) is implicit, because the relevant opacity $\kappa$ depends on $T$ and $\tau$.  To be specific, for the optically thin and thick limits we use the Planck and Rosseland means, $\kappa_{\rm Pl}$ and $\kappa_R$, respectively.   Numerically, we adopt the values computed by \citet{2003A&A...410..611S} for composite-aggregate grains.  We fit $\kappa_{\rm Pl} \simeq 6.1\deltagr (T/100\,{\rm K})^{1.58}$\,cm$^2$\,g$^{-1}$ and  $\kappa_R \simeq 3.0\deltagr  (T/100\,{\rm K})^{1.93}$\,cm$^2$\,g$^{-1}$, respectively, for $T<150$\,K, and $(\kappa_{\rm Pl},\kappa_R) \simeq (11.6,6.6)\deltagr$\,cm$^2$\,g$^{-1}$ for $150<T<400$\,K.  The relative grain abundance $\deltagr$ reflects the metallicity  in the regime of interest: $\deltagr\simeq Z/Z_\odot$.   We  assume $T<150$\,K for most of this work, but consider higher temperatures in \S\,\ref{S:Discussion}. 

With these evaluations, equation (\ref{eq:Tapprox}) is an interpretable approximation to the multi-frequency radiation transport solution, for which \cite{2005ApJ...631..792C} present more accurate and complicated solutions in spherical symmetry.  Equation (\ref{eq:Tapprox}) ignores the fact that the radiation field tends to reflect higher temperatures than the local dust temperature, which makes our evaluation of $\kappa$ an underestimate.  It also neglects the tendency of radiation to flow around dense regions within a clumpy dust distribution.  
These corrections are likely to act in opposite ways.    
  
For the luminosity $L$ we consider three scenarios, avoiding the regime in which rapid accretion causes the star to swell in size \citep{1990ApJ...360L..47P}. 
First, for an individual low-mass star with $m_*<M_\odot$ accreting at a rate $\dot m_*$, we are interested in the low-$\dot m_*$ regime in which deuterium burning sets the gravitational potential, so that $L_* \simeq \Psi_{*D} \dot m_*$.  
Second, for an individual massive star with $M>10M_\odot$ we assume $L_*$ limits to the zero-age main sequence (ZAMS) luminosity.  We can either evaluate this from stellar models \citep[e.g.,][]{1996MNRAS.281..257T}  or approximate it using the radiative diffusion luminosity in Eddington's standard model, in which   
\begin{equation} \label{eq:EddingtonStandardModel} 
\frac{\Gamma_*^{1/2}}{(1-\Gamma_*)^2 } = \frac{2^{3/2} m_*}{m_{*,\rm rad}} \end{equation} 
 implying $\Gamma_*\rightarrow 1$ for $m_*\gg m_{*,\rm rad}$. Here
 \begin{equation} \Gamma =\frac{L}{L_{\rm edd}} =  \frac{\kappa L }{4\pi G M c} \end{equation} 
 is the Eddington ratio and $\Gamma_*$ is its value within a massive star, where $(L,M,\kappa) =(L_*,m_*,\kappa_{\rm es})$.   
Finally, for a stellar population with total stellar mass $M_* = f_* M$, assuming none of the stars have evolved,
 we consider $(L/M)_*$ to be of order the IMF-averaged ZAMs value.  
 
The Eddington factor of the cluster-forming cloud is 
\begin{equation} \label{eq:GammaEdd}
\Gamma =  f_* \left(\frac{L}{M}\right)_{*,3} \frac{\kappaCGS}{13} 
\end{equation} 
where $(L/M)_* = 10^3 (L/M)_{*,3}\, L_\odot/M_\odot$ and `cgs' means evaluated in those units, so $\kappa = \kappaCGS$\,cm$^2$\,g$^{-1}$ and $\Sigma = \Sigma_{\rm cgs}$\,g\,cm$^{-2}$.    
Evaluating equation (\ref{eq:Tapprox}), our choices imply a characteristic temperature for the entire region, 
\begin{eqnarray}\label{eq:T}
T &\simeq&  61 \,{\rm K} \times \left\{ 
\begin{array}{lcc} 
[\Gamma/(\deltagr^2 f_g)] ^{0.14} & ~~~~& \tau <1 \\
\Gamma^{1/4}f_g^{1/4} \SigmaCGS^{1/2} & ~~~~ & \tau >1
 \end{array}
\right.
 \\  
 &\rightarrow& \left\{  \nonumber
\begin{array}{lcc} 
 47 [f_* (L/M)_{*,3}/(f_g \deltagr) ]^{0.18}\,{\rm K} & ~& \tau <1 \\
 58[f_* f_g  \deltagr  (L/M)_{*,3}]^{0.48} (\Sigma_{\rm cgs}/3)^{0.96} \,{\rm K}, &~ &  \tau >1
 \end{array}
\right.
\end{eqnarray} 
so long as these give $T<150$\,K (for the fits to $\kappa$ to be valid) and $\Gamma\lesssim 1$ (for dynamical stability). 
 
With this temperature, the self-consistent optical depth  can be written $\tau \simeq \left\{ \Sigma/\Sigma_{\rm tr}, (\Sigma/\Sigma_{\rm tr})^{2.86}\right\}$ for optically \{thin, thick\} regions, where in each regime, 
\begin{equation}\label{eq:tau} 
\Sigma_{\rm tr} = 
 \left\{  \nonumber
\begin{array}{lcc} 
 \frac{1.1 \,\rm g\,cm^{-2} }{ [ (f_g/0.5)\deltagr]^{0.71} \left[(f_*/0.5) (L/M)_{*,3}\right]^{0.28}} & ~& \tau <1 \\
\frac{ 4.0\,\rm g\,cm^{-2} }{[ (f_g/0.5) \deltagr]^{0.67} \left[(f_*/0.5) (L/M)_{*,3}\right]^{0.32}}  &~ &  \tau >1
 \end{array}
\right.
\end{equation} 
From this, we see that $(L/M)_{*,3}^{0.3} \deltagr^{0.7} \SigmaCGS\simeq 2$  divides optically thick and thin regions halfway through the star formation process (when $f_g=f_*=0.5$), although the transition is not sharp: it extends a factor of two on either side of this boundary. 

The Eddington factor $\Gamma$ provides an important dynamical limit on the luminosity-to-mass ratio, as explored by \cite{2018MNRAS.481.4895C}.   
It represents the ratio of between the outward force of radiation and the inward force of gravity on the dusty gas; in the optically thin case, this is equivalent to the ratio of radiation and hydrostatic pressure gradients. 
Gas is rapidly expelled when $\Gamma >1$.   In fact, dynamically-important magnetic fields are likely to reduce the critical value of $\Gamma$, while inhomogeneities may allow matter to persist in high-column concentrations even as low-column regions are blown away.  

To actually achieve radiation-driven blowout from an optically thick cloud, a top-heavy IMF is required.    In a normal \cite{2001MNRAS.322..231K} or \cite{2005ASSL..327...41C} IMF with a $120\,M_\odot$ cutoff, the ZAMS luminosity is $(L/M)_{*,3}\simeq 1$; and as $f_*<1$ and $\kappa_{R,\rm cgs}\lesssim 6.6\deltagr$\,cm$^2$\,g$^{-1}$, this implies $\Gamma\leq 0.51$ in equation (\ref{eq:GammaEdd}). 
  On the other hand, top-heavy IMFs are significantly more luminous; changing (only) the high-mass slope to $\alpha=(2.0,1.8)$ leads to $(L/M)_{*,3}\simeq (2.3,3.3)$.  Increasing the peak of the IMF also leads to a more luminous population, as does raising the upper mass limit (especially if $\alpha<2$).   Individual stars  have have $(L/M)_{*,3} \simeq  65\Gamma_*/(1+X)$ with $\Gamma_*\gtrsim 0.5$ for $m>m_{*\rm rad}$, which means that sufficiently top-heavy young populations can achieve $\Gamma>1$.   (Note that  \citealt{2024ApJ...967L..28M} have recently studied the consequences of a top-heavy IMF for stellar feedback.) 
 
In the extrapolation of non-thermal turbulent velocities from the cloud scale to the sonic length, we adopt the standard scaling $\sigma_{\rm NT}(r) = \sigma_c (r/R_c)^\eta$ and assume $\eta\simeq\frac12$ in the supersonic regime (Mach number ${\cal M}\geq 1$).
This is a simplifying idealization of what is actually a complicated and intermittent turbulent velocity field.  
 
A final ingredient is the critical state of a thermally supported object.   For an isothermal sphere the critical mass per unit radius is $2.4c_s^2/G$ \citep{1955ZA.....37..217E,1956MNRAS.116..351B}; likewise, for an isothermal filament the critical mass per unit length is $2 c_s^2/G$ \citep{1964ApJ...140.1056O}, where $c_s = \sqrt{kT/\mu}$ is the isothermal sound speed and $\mu=2.3m_p$ is the mean molecular weight of molecular material.  When non-thermal support is relevant, we adjust these formulae by replacing $c_s$ with an effective sound speed $c_{\rm eff}$ that ideally accounts for turbulent and magnetic energy.  For simplicity, in evaluating critical masses we ignore the energy in static and perturbed \citep{1995ApJ...440..686M} magnetic fields, adopting $c_{\rm eff}^2 \simeq \sigma^2  = c_s^2 + \sigma_{\rm NT}^2$ where $\sigma$ is  the total line width. 

\section{Opacity-limited fragment mass} \label{S:Mmin} 
During collapse and fragmentation, the minimum  mass of a fragment is the so-called opacity limit \citep{1976MNRAS.176..367L,1976MNRAS.176..483R}: 
\begin{eqnarray}\label{eq:Mmin} 
m_{\rm opac} &\simeq & \left( \frac{ c_s}{f_r c}\right)^{1/2}  \frac{\mPlanck^3}{\mu^2}  
\\ \rightarrow&& 
0.073M_\odot  \left\{\begin{array}{lcc} 
\deltagr^{0.09} [ \left({L}/{M}\right)_{*,3}f_*/f_g]^{0.24} && \tau<1 \\ 
\deltagr^{1.0} [\left({L}/{M}\right)_{*,3}f_* f_g]^{0.71} \left(\frac{\SigmaCGS}{3}\right)^{1.43}, && \tau > 1.  
\end{array} \right.  \nonumber
\end{eqnarray} 

%{\mbox{~}}

\noindent where $f_r\leq 1$ is the radiative efficiency, which is determined by radiation diffusion in the regime of interest.  Accordingly we set $f_r = 1/q\tau_{\rm opac}$ where $\tau_{\rm opac}=\kappa(T) m_{\rm opac}/\pi r_{\rm opac}^2$ is the self-consistent optical depth.     In the evaluation  we set  $q^{1/3}=13.5$ to match equation (26) of \cite{1999ApJ...510..822M}.

\section{Sonic mass}\label{S:Msonic} 

The peak of the IMF is likely to originate in marginally stable objects at the sonic scale: those with initial radius $r_s = [c_s/\sigma_{\rm NT}(R)]^{1/\eta} R$, the value for which turbulent motions match the thermal sound speed \citep{2003ApJ...585L.131V,10.1093/mnras/stv872}.  For a sonic core $\sigma_{\rm NT}(r_s) = c_s$ so that  $c_{\rm eff}^2 \simeq 2 c_s^2$; the core mass is therefore
\begin{equation} \label{eq:ms_general} 
m_s \simeq 4.8\frac{c_s^2}{G} r_s= 0.96 \frac{\alphavir}{\Mach^{2+1/\eta}} M = 7.6 \frac{c_s^4 {\cal M}^{2-1/\eta} }{\alphavir G^2\Sigma}
\end{equation}   where $\Mach=\sigma_{\rm NT}/c_s$ is the Mach number of the environment, evaluated at the local sound speed $c_s$.   We assume $\eta=1/2$ for the rest of this section, but other values can be accommodated by applying the factor ${\cal M}^{1/\eta-2}$.  Using equation (\ref{eq:Tapprox}) \newtext{along with the definition of $\Gamma$}, we find that 
\begin{eqnarray}  \label{eq:Msonic_Env}
m_{s} &\simeq& 12.6 \frac{\sqrt{f_g\Gamma}}{ \alphavir} \frac{\mPlanck^3}{\mu^2}\times
\left\{ 
 \begin{array}{lcc}
2.66/\tau,&~& \tau<1 \\
1&~& \tau >1. 
\end{array} \right. 
\end{eqnarray}

We see that when a cloud is in a particular state -- when it is (1) translucent or optically thick  ($\tau\gtrsim 1$), (2) at or near the Eddington limit ($\Gamma\lesssim 1$), and (3) suffused with turbulence at or near the spectral slope of Burgers turbulence ($\eta\simeq1/2$) -- all three of which  are likely to be satisfied during intense episodes of massive star cluster formation  -- then its sonic mass will be anchored to a few times the fundamental mass scale for molecular gas, 
\[ \frac{ \mPlanck^3}{\mu^2}= 0.34 M_\odot.\]  This is a primary conclusion of our work.    

To be  specific, we note that for $\Gamma=\alphavir=1$, $m_s \rightarrow 3.1 ({f_g/0.5})^{1/2}\,M_\odot$ in the optically thick limit.  We expect $\Gamma$ to saturate at a value less than unity, either because of dynamical effects other than radiation pressure, or because of roughly constant opacity for temperatures above 150\,K (see \S\,\ref{S:Discussion}), so the maximal value of  $m_s$ in the optically thick regime is likely to be less than $\sim3\,M_\odot$. 

 With $\eta=1/2$, thermal cores have a column density comparable to their environment ($\Sigma_s = m_s/\pi r_s^2 = 0.96\alphavir \Sigma$).  Their optical depths are therefore comparable (if the dust temperature matches the environment) or somewhat greater  (if they are warmed from within by a forming star) to that of the environment. 
   
These two possibilities imply two rules for the sonic mass. 
If the temperature is set by the environment (and $T<150$\,K) one finds, 
\begin{equation} \label{eq:ms_env}
m_{s,{\rm env}}\simeq
 \left\{ 
 \begin{array}{lcc}
\frac{0.78}{\alphavir} \left(\frac{\SigmaCGS}{3}\right)^{-1} \left[\frac{f_* (L/M)_{*,3}}{f_g \deltagr} \right]^{0.36}\, M_\odot & \tau<1 \\
\frac{1.22}{\alphavir} \left(\frac{\SigmaCGS}{3}\right)^{0.93}  \left[ f_*f_g \deltagr (L/M)_{*,3} \right]^{0.97}\, M_\odot & \tau >1 
\end{array} \right.
\end{equation} 
where all the parameters refer to the parent cloud.   (In formulae like this we assume $T<150$\,K and we do not explicitly impose $\Gamma<1$; these are additional constraints.) 

On the other hand, if the thermal core is heated\footnote{\cite{2020ApJ...904..194H} argue that the protostellar luminosity problem \citep{1990AJ.....99..869K} is evidence that protostellar heating is ineffective.  However, \cite{2011ApJ...736...53O} find that accelerating and unsteady accretion suffices to resolve the problem, which leaves open the possibility that a sonic core can regulate its own mass. }   by the luminosity of an accreting protostar ($L_*\simeq \Psi_{*D} \dot{m}_*$),
then its mass is tied to the fundamental constants that set $\Psi_{*D}$ \citep{2011ApJ...743..110K}.  One must determine the local value of  $c_s$  self-consistently with the accretion rate $\dot{m}_*\simeq c_{\rm eff}^3/G=2^{3/2}c_s^3/G$.   Doing this, we obtain 
\begin{equation} \label{eq:ms_self} 
m_{s,{\rm self}} \simeq 
 \left\{ 
 \begin{array}{lcc}
0.56  \deltagr^{-0.33} [\alphavir (\SigmaCGS/3)]^{-0.67} \,M_\odot &~& \tau_s<1 \\
0.47\deltagr^{0.78} [\alphavir (\SigmaCGS/3)]^{1.33}\,M_\odot  &~& \tau_s >1 
\end{array} \right.
\end{equation} 
where $\tau_s$ is the optical depth of the self-irradiated core; self-consistently $\tau_s>1$ when $\alphavir \Sigma_{\rm cgs} \deltagr^{1/2} \gtrsim 4.2$.   
Comparing this estimate of $m_{s,{\rm self}}$ to the environmentally-heated value $m_{s,{\rm env}}$, and accounting for details like the fact that optically-thin cores can only exist within optically-thin environments, suggests that self-heating by accretion luminosity is important either for environments that are not particularly luminous ($f_* (L/M)_{*,3}<1$) or for regions of high column density in which the core is optically thick to its accretion luminosity.  

However, the last conclusion is unlikely to be valid because there is (quite literally) a hole in the logic leading to the optically thick branch of equation (\ref{eq:ms_self}): it follows from the diffusion approximation in spherical symmetry, and does not account for a protostellar outflow cavity.   An outflow cavity strongly modifies the problem of radiation diffusion in the thick limit, by providing an optically thin escape path for protostellar radiation \citep{krumholz2004protostellar,2005ApJ...628..817M}.    While detailed analysis is in order (along the lines of \citealt{2017A&A...600A..11R}), we can roughly account for this effect by setting $f(\tau)\simeq 1$ in  equation (\ref{eq:Tapprox}).  This gives the modified optically thick solution 
\begin{equation}\label{eq:ms_self_ThickWithOF} 
m_{s,{\rm self}} \simeq   \begin{array}{lcc}
1.0 [\alphavir (\SigmaCGS/3)]^{-1/9} M_\odot &~& \tau_s >1 
\end{array}
\end{equation} 
which both highly insensitive to the environment, and consistently lower than the result for a core heated by an optically-thick environment (equation \ref{eq:ms_env}) -- except for especially low values of the combination $f_g g_* \deltagr \Sigma (L/M)_*$, which are unlikely to be consistent with $\tau>1$ in any case.  

Based on this, we tentatively conclude that the sonic mass scale is self-consistently determined by the surrounding radiation environment in the context of intense stellar cluster formation, whereas internal accretion luminosity is the relevant heating source in contexts more like well-studied Galactic environments. 
 
 \section{Upper mass scale} \label{S:Mupper} 
 
Between $m_s$ and the cloud mass $M$, the only physical effect clearly associated with a mass scale has to do with the stellar luminosity: $\Gamma_*$  approaches unity as  $m_*$ exceeds $m_{*,\rm rad}=140\,M_\odot$.  Note that an optically thick dusty region that is bound by the gravity of a single massive star has $\Gamma_{\rm cl} \simeq 13\deltagr\Gamma_*$. Using \autoref{eq:EddingtonStandardModel}, this means that some critical value of $\Gamma_{\rm cl}$ (such as unity) can be associated with a value of $m_*$ so long as $\deltagr\gtrsim \Gamma_{\rm cl, \rm crit}/13$.  The stellar upper mass limit is likely to be related to $m_{*,\rm rad}$ either as a result of the stars' own stability and wind production (due to high Eddington factors and radiation pressure support), or due to the super-Eddington nature of stellar accretion in dusty gas (if the metallicity is not too low).   Indeed, numerical experiments with sufficiently low metallicity to achieve $\Gamma_{\rm cl}<1$ around very massive stars do find that these continue to accrete well above $m_{*\rm rad}$ (M.\,Grudi\'c, private communication, 2024). 

\subsection{Stabilization of massive-star accretion flows} \label{SS:Mupper_stab} 

Massive stars (or star systems) grow from regions that encompass many sonic masses, which makes them sensitive to the stability of the regions that feed them.  The more stable the feeding zone, the more successfully a growing massive star can  accumulate gas (rather than companions).  Our analysis of this will be similar to that of \citet{2008Natur.451.1082K}, except that we assume the region is significantly heated by starlight. 

As accretion tends to occur along dense filaments, and as the critical mass per unit length $\lambda$ of a thermally-supported filament is $\lambda_{\rm th}= 2c_s^2/G$, we conclude that the character of the accretion flow is determined by the stability parameter 
\begin{eqnarray} 
\xifil
& = &  \frac{2 c_s(r)^2 r}{G M(r)} = \frac25\frac{\alphavir(r)} {{\cal M}(r)^{2}},
 \end{eqnarray} 
 which we can evaluate using equation (\ref{eq:ms_general}) as 
 \begin{eqnarray}\label{eq:xiFil_vs_MonMs} 
 \xifil &=& \frac25\alphavir \left( \frac{m_s}{0.96 \alphavir M}\right)^{\frac{2\eta}{2\eta+1}}  
 \\ \nonumber &&\xrightarrow[]{\eta=1/2}  \left( \frac{\alphavir}{6.0} \frac{m_s}{M} \right)^{1/2} .
 \end{eqnarray} 

We envision that the collapsing accretion zone, with mass distribution $M(r)$, contains a range of filamentary structures with values of $\lambda$ distributed around the characteristic value $M(r)/r$ in a manner determined by the physical effects at play \citep{2017MNRAS.465.2254K} -- and that the fraction of this region capable of accreting before it undergoes fragmentation depends on the combination $\lambda_{\rm th} r/M(r) = \xifil$.   If  accretion involves a number of filaments $\Nfil$ that exceeds $1/\xifil$,  it is possible for the accretion flow to be completely stabilized by thermal pressure alone (see \citealt{2015ApJ...815...68M}).  

Turbulent motions, magnetic fields \citep{2000MNRAS.311..105F}, and  tidal gravity  \citep{2018A&A...611A..89L} also stabilize the region.  Unlike these,  the thermal support   parameterized by $\xifil$ is clearly a function of the star formation activity and radiative transfer, rather than dimensionless properties of the initial conditions.  (The same might be said for non-ideal magnetic effects, but these lie beyond the scope of our work.) 

Equation (\ref{eq:xiFil_vs_MonMs}) indicates that the stability of an accretion zone depends on the number of thermal core masses it contains, which in turn reflects the radiation environment.   Effects that increase $m_s$ also enhance the stability of accretion.     And, just as a thermal core can be self- or environmentally-heated, a massive star's accretion zone can be heated and stabilized either by the star itself, or by the radiation environment of its parent region.   We stress that the accretion flow can feed a growing stellar system or sub-cluster as opposed to a single star (\citealt{2006MNRAS.373.1563K}, \citealt{2010ApJ...708.1585K}, \citealt{2015ApJ...815...68M}), and stabilization provides a path to flatten the high-mass slope of the IMF as seen in simulations by \cite{2011ApJ...740...74K}, \cite{2011ApJ...742L...9C}, and \cite{2013ApJ...766...97M}.  
It also provides an explanation for primordial mass segregation in optically-thick cluster environments; we note that  \cite{2008ApJ...685..247B} present evidence for primordial segregation in globular clusters. 
We return to this point in \S\,\ref{SS:alpha_high}. 

\section{Dynamical influence of protostellar outflows} \label{S:outflows}
We previously considered the possibility that protostellar outflows will affect the ability of accreting protostars to set their own accretion reservoir.  Now let us consider the role of protostellar winds and jets in returning matter to the environment, a process that limits the efficiency of individual or clustered star formation \citep[][hereafter MM00]{2000ApJ...545..364M}. 
We assume that the outflow momentum per stellar mass, denoted $f_w v_w$ for consistency with MM00, is related to the stellar Keplerian speed $\Psi_*^{1/2}$ by a factor $\phi_w\simeq 0.1-0.2$ that accounts for the ratio of wind to accreted matter as well as the ratio of wind speed to Keplerian speed.  That is, 
\begin{equation} 
f_w v_w = \phi_w \Psi_*^{1/2}. 
\end{equation} 
Using this, we can obtain MM00's parameter $X$ (their eq.\,3, in which $c_g \ln 2/\theta_0\simeq 5.8$) and the single-star formation efficiency $\varepsilon$ (their eq.\,31, for spherical cores, or eq.\,44, for flattened cores) or larger regions (their eq.\,55). 

To evaluate $\varepsilon$ for thermal cores of the sonic mass, we take $\Psi_*^{1/2}=\Psi_{*D}^{1/2}= 195$\,km\,s$^{-1}$, and we ignore the possibility that the core is flattened (or more likely, elongated but misaligned from the outflow axis).   The  escape velocity of such a core is $v_{{\rm esc},s} \simeq 3.1 c_s$, according to the rule $m_s/r_s \simeq 4.8 c_s^2/G$.    For the case of an environmentally-heated core we can evaluate $c_s$ from equation (\ref{eq:T}) to obtain 
\begin{equation} \label{eq:efficX_senv}
X \simeq \frac{0.44}{\phi_w/0.1} \times 
\left\{ 
 \begin{array}{lcc}
(\Gamma/f_g)^{0.07} \deltagr^{-0.14} &~& \tau<1 \\
(\Gamma f_g \SigmaCGS^2)^{1/8} &~& \tau >1 
\end{array} \right.
\end{equation} 
 for stars forming within regions of mass $m_{s,{\rm env}}$, from which (neglecting the small wind mass correction in MM00 eq.\,31)
\begin{equation}\label{eq:epsilon} 
\varepsilon = \frac{2X}{1 + \sqrt{1 + (2X)^2} }\simeq X(1-X^2)
\end{equation} 
where the approximation is valid to 10\% for $X<0.5$. 

 We see that environmentally-heated thermal cores form stars at about $38\%$ efficiency according to this estimate.  (It is possible but not certain that smaller cores, even down to $m_{\rm opac}$, might develop outflows and form at a similar efficiency.)  Self-heated cores would form at moderately higher efficiency, thanks to their elevated temperatures.

For massive star formation, a couple factors are different.  First, the stellar potential well of a massive star is deeper.
Second, the relevant escape velocity reflects the entire accretion zone, which might feed more than a single object, and this region is not thermally supported.   Together, we find that 
\begin{equation}\label{eq:efficX_massive}
X \simeq \frac{0.42}{\phi_w/0.1}
 \frac{\Psi_{*H}^{1/2}}{\Psi_*^{1/2}} \left(\frac{M}{m_{*,\rm rad}} \SigmaCGS\right)^{1/4} 
\end{equation} 
\noindent for massive stars within their accretion regions (normalizing $M$ to the stellar mass scale $m_{*\rm rad}=140M_\odot$, and $\Psi_*$ to the value defined earlier for H-burning convective stars). This indicates that individual massive stars can remove a significant amount of matter from their accretion environments, and form with an outflow-limited efficiency very similar to that of individual low-mass stars. 

There may exist stars that form at higher efficiency, because of the reduction in $\Psi_*$ associated with an inflated protostellar radius, but we do not attempt to calculate this effect.  

\section{Influence of binary and multiple formation} \label{S:binarity} 
Another key factor affecting the numbers and masses of stars, relative to a thermal scale such as $m_s$, is the fragmentation into binary or multiple star systems.  This is obviously important for the lower IMF, insofar as heat from starlight and viscous dissipation is critical to stifling the fragmentation of  protostellar disks  into large numbers of brown dwarfs \citep{2005ApJ...628..817M,2009MNRAS.392.1363B,2009MNRAS.400.1563S,2009ApJ...703..131O,2010ApJ...710.1343U}.  At somewhat higher masses, fragmentation will tend  to suppress the IMF peak mass by some factor $\phi_{\rm mult}\sim 0.5$, as the observed companion frequency is of order 0.6 around $1M_\odot$ (and strongly correlated with the mass of the primary, at least for local stellar populations; see \citealt{2023ASPC..534..275O}). 

We wish to infer how $\phi_{\rm mult}$ is likely to depend on environmental factors. 
Our strategy will be to evaluate the specific angular momentum $j_s$ of a sonic core, assume some fraction of this is not lost to magnetic braking during star formation, and then examine the gravitational stability of the structure (such as an accretion disk) that results.  
In this we follow \citet{2005ApJ...628..817M}, \citet{2006MNRAS.373.1563K}, and \cite{2010ApJ...708.1585K}, hereafter ML05, KM06, and K+10, respectively.  Our procedure
is designed to capture close binary formation from disk fragmentation, rather than the more stochastic turbulent channel for the formation of wide binaries that is inferred both from simulations \citep{2010ApJ...725.1485O} and observations \citep{2019ApJ...875...61M,2019MNRAS.482L.139E}.  

For a sonic core  $j_s = \theta_j r_s \sigma_{NT}(r_s) = \theta_j r_s c_s$, where we consider $\theta_j$ to be chosen from a Maxwellian distribution peaked around $0.383(\eta/0.5)^{0.49}$. In this we follow KM06's evaluation of the angular momentum contained within an uncorrelated Gaussian-random velocity field with line width-size index $\eta$, when integrated over the density structure of a critical Bonnor-Ebert sphere in the manner of \cite{2000ApJ...543..822B}.   Assuming  a fraction $f_j$ of the angular momentum survives magnetic braking to form stellar system with mass $m_s/\varepsilon = 4.8 c_s^2 r_s/\varepsilon G$ this implies an orbital radius of about $r_s f_j^2/33\varepsilon$, or 
\begin{eqnarray} \label{eq:a_orb} 
a_{{\rm orb},s} &\simeq& 107  \frac{f_j^2}{\varepsilon \alphavir}{\rm AU} \times 
\left\{ 
 \begin{array}{lcc}
[\Gamma/(\deltagr f_g)]^{0.14} \SigmaCGS^{-1} &~& \tau<1 \\
\left[f_g\Gamma\right]^{1/4} \SigmaCGS^{-1/2} &~& \tau >1 
\end{array} \right.
\\ \nonumber  &\rightarrow& 
\left\{ 
 \begin{array}{lcc}
 80 \frac{f_j^2}{\varepsilon \alphavir } [f_* (L/M)_{*,3}/(\deltagr f_g) ]^{0.18} \SigmaCGS^{-1}\,{\rm AU}&~& \tau<1 \\
35 \frac{f_j^2}{\varepsilon \alphavir } \left[\deltagr f_g f_*  (L/M)_{*,3}\right]^{0.48} \SigmaCGS^{-0.03}  \,{\rm AU}
 &~& \tau >1 
\end{array} \right.
\end{eqnarray} 
for environmentally-heated cores. 
 
 From the system accretion rate $\dot M_{\rm sys}$, system mass $M_{\rm sys}$,  specific angular momentum $j$, and local (e.g., disk) sound speed $c_{sd}$ one can compute the dimensionless parameters 
 \[ \xi_d = \frac{G \dot M_{\rm sys} }{ c_{sd}^3} ~~ {\rm and} ~~ \Gamma_d = \frac{j^3 \dot M_{\rm sys} }{G^2 M_{\rm sys}^3}. \] 
 A disk that forms with these parameters has a characteristic aspect ratio of order $(\Gamma_d/\xi_d)^{1/3}$, as discussed by K+10.  In the thin-disk limit $(\Gamma_d/\xi_d)^{1/3}\ll1$,  \citeauthor{1964ApJ...139.1217T}'s (\citeyear{1964ApJ...139.1217T}) stability parameter is $Q_d\simeq 3\alpha_d/\xi$ \citep{2001ApJ...553..174G}, where $\alpha_d$ is an effective viscosity parameter meant to capture transport by the gravitational instability. If we adopt ML05's  estimate for the maximal viscosity parameter (which is $ \max \alpha_d=0.23$), we find that sufficiently thin disks should fragment when $\xi\gtrsim 1.12$.  However,  K+10 show that finite thickness stabilizes accretion, and identify a fragmentation boundary $\xi_d \gtrsim (850\Gamma_d)^{2/5}$ at the finite resolution of their simulations.  To be definite, we take
  \begin{equation} \xi_d>\max[1.12,(850\Gamma_d)^{2/5}] \end{equation} 
  as the fragmentation criterion, while acknowledging that both branches of this formula are uncertain.  
  
 Applied to a sonic core for which $\dot M_{\rm sys} = \varepsilon c_{\rm eff}^3/G = 2^{3/2} \varepsilon c_s^3/G$, $M_{\rm sys} = \varepsilon m_s$, and $j = j_s$, we find $\Gamma_d\simeq (\eta/0.5)^{1.47}/(700 \varepsilon^2)$ so that $(850 \Gamma_d)^{2/5} = 2.25 (\varepsilon/0.4)^{-4/5}$.      As this value exceeds 1.12, we take \[ \xi_{d,{\rm crit},s} = 2.25 (\varepsilon/0.4)^{-4/5}\] to be the critical value of $\xi_d$ for a sonic core (adopting $\eta=0.5$ at this point, for simplicity).  We note that this  implies that  fragmentation will occur when the outer disk becomes marginally colder than its parent core. 
 
To determine whether disks do fragment by this criterion, we must compute $\xi_d$. We assume the outer disk is optically thick; this is reasonable, as the disk column density $\Sigma_d$ is self-consistently 1.6 to 2 dex higher than that of its parent core, which in turn is comparable to that of the environment.   

First, consider a disk warmed by its own internal dissipation, rather than by starlight. Using the logic leading to ML05 eq.~(31),  we see that the value of $\xi_d$ in such a disk is directly related to its outer orbital frequency.  We equate the local  cooling rate $4 a c T_d^4/(3\kappa_d\Sigma_d)$ with the heating rate $3\dot{M}_{\rm sys} \Omega_d^2/8\pi$, in the context of viscous accretion at the rate $\dot{M}_{\rm sys} = 3\pi \alpha_d \Sigma_d c_d^2/\Omega_d$.  Solving for $\xi_d = G\dot{M}_{\rm sys}/c_d^3$ and naming the result $\xi_{d,{\rm visc}}$, we find  
\begin{eqnarray} \label{eq:xid}
\xi_{d,{\rm visc}} = \left[\frac{2 \alpha_d a c}{3 (\kappa_d/T_d^2)}\right]^{1/2}  \frac{4\pi G \mu}{k} \Omega_d^{-3/2} = \left(\frac{\Omega_{\xi}}{\Omega_d}\right)^{3/2} 
\end{eqnarray} 
where $\Omega_d$ is the orbital frequency of the outer disk, and the period corresponding to $\xi_{d,{\rm visc}}=1$ is 
\begin{equation} \label{eq:Omega_xi} 
\frac{2\pi}{\Omega_{\xi} }  \simeq 500 \left(\frac{\deltagr}{\alpha_d/0.23}\right)^{1/3}\,{\rm yr}. 
\end{equation} 
In this evaluation we assume that the opacity law is unchanged between the core and the disk, so that $\kappa_d/T_d^2 \simeq 3.0\times10^{-4}\deltagr$\,cm$^2$\,g$^{-1}$ at the relevant temperatures.    

Using equation (\ref{eq:xid}), our fragmentation criterion corresponds to disk periods $\gtrsim 860(0.23\deltagr/\alpha_d)^{1/3}(\eta/0.5)^{0.37}(\varepsilon/0.4)^{-0.53}$\,yr, whereas the typical disk periods are a few times longer if all parameters ($f_j$, $\SigmaCGS$, etc.) are taken to be unity.  Phrased in terms of the stability parameter $\xi_d$, the comparison implies 
\begin{eqnarray} \label{eq:ViscDiskStability}
\frac{\xi_{d,{\rm visc}} }{\xi_{d,{\rm crit},s}} &\simeq 7.9 \frac{f_j^{9/2} (\alpha_d/0.23)^{1/2}}{\alpha^{3/2} (\varepsilon/0.4)^{11/5} } \times 
\left\{
 \begin{array}{lcc}
\frac{(\Gamma/f_g)^{0.10} }{\deltagr^{0.71} \SigmaCGS^{3/2} } & ~& \tau<1\\ 
\frac{ (f_g \Gamma)^{3/16}}{\deltagr^{1/2}  \SigmaCGS^{9/8} }&~&\tau>1
 \end{array} 
  \right.  
  \\ \nonumber& \rightarrow 
1.3 \frac{  f_j^{9/2}  (\alpha_d/0.23) }{\alpha^{3/2}(\varepsilon/0.4)^{11/5}}  \times 
 \left\{
 \begin{array}{lcc}
\frac{ [f_* (L/M)_{*,3}/f_g]^{0.13}}{\deltagr^{0.63} (\SigmaCGS/3)^{3/2} }& ~& \tau<1\\ 
\frac{ [f_g f_* (L/M)_{*,3} ]^{0.36}}{\deltagr^{0.14} (\SigmaCGS/3)^{0.78} }.& ~& \tau>1
 \end{array} 
  \right.  
\end{eqnarray} 
 We see that  for fiducial parameters, viscous dissipation will not prevent the disk formed from an environmentally-heated sonic core from fragmenting.   The removal of mass by protostellar outflows  ($\varepsilon$) is  a destabilizing factor, as it increases $a_{\rm orb}$ and decreases $\Omega_d$.  However, it would take only a mild amount ($f_j\lesssim 0.63$) of magnetic braking to overcome this effect, and make viscous heating more of a barrier to fragmentation.    

Considering environmental parameters, we see that fragmentation is suppressed  by an increase in the column density of the environment, as cores with higher $\Sigma$ make smaller and more stable disks.   Increasing the metallicity also suppresses fragmentation (through $\deltagr$), primarily because the disk is optically thick.  (This implies a metallicity/close-binary anticorrelation like the one discovered by  \citealt{2019ApJ...875...61M}.)  

According to equation (\ref{eq:ViscDiskStability}), disks in more luminous regions are less stable because this increases $r_s$ and hence $a_{\rm orb}$.  However, the stabilizing influence of irradiation is not yet included; we turn to this next. 

To consider the case where the irradiating source is the star, with $L_* = \Psi_{*D} \dot M_{\rm sys}$, we adopt the model developed by ML05 in which starlight strikes the inner edge of an outflow cavity and then indirectly warms the disk to an equilibrium temperature determined by $a c T_{d,{\rm irr}} ^4 \simeq 0.1 \varepsilon^{0.35} L_*/( \pi a_{\rm orb}^2)$.   This leads to a moderate suppression of fragmentation, in that 
\begin{eqnarray}  \nonumber
\frac{\xi_{d,{\rm irr}}} {\xi_{d,{\rm crit},s}} \simeq 0.63 \frac{ f_j^{3/2} } {\alpha^{3/4} } 
\left(\frac{\varepsilon}{0.4}\right)^{0.81}
 \times 
\left\{
 \begin{array}{lcc}
\frac{(\Gamma/f_g)^{0.24} }{\deltagr^{0.47} \SigmaCGS^{3/4} } & ~& \tau<1\\ 
 (f_g \Gamma)^{0.42} \SigmaCGS^{0.09}. &~&\tau>1
 \end{array} 
  \right.  \\
\end{eqnarray} 
Even in the absence of stellar luminosity, a disk will be warmed by its environment.  When the environment is optically thick, this implies $c_d\gtrsim c_s$ and therefore $ \xi_{d,{\rm irr}}/\xi_{d,{\rm crit},s} \lesssim 0.5 (\varepsilon/0.4)^{9/5} $. (In the optically thin case, environmental irradiation is less stabilizing when the environment is not as luminous; additional factors like $\SigmaCGS^{-3/8}$ appear.) 

Although we have seen that stellar and environmental irradiation are stabilizing factors, especially for optically thick cores, these estimates are close enough to unity that we do not expect them to entirely shut off binary and multiple star formation on their own.  As \citet{2023ASPC..534..275O} stress, fluctuations in the mass accretion rate and central luminosity are likely to provide a route for fragmentation. 

Briefly considering self-heated  cores, we note that heating by the central object makes a sonic core larger, which increases $a_{\rm orb}$, which increases the likelihood of fragmentation.  As discussed in \S\,\ref{S:Msonic}, however, the presence of an outflow cavity is likely to prevent sonic cores from being self-heated in the contexts of interest. 

\section{Discussion}\label{S:Discussion} 

We have found that the mass scales and modifying factors that affect the stellar mass distribution all depend to some extent on the  parameters of the region, but that the dependence is not necessarily monotonic because of the shift from optically thin to optically thick behavior.    Relevant factors include the column density and dust content, as well as the light-to-mass ratio of the stellar population.  It is also relevant that the Eddington ratio cannot become large, as this fact connects the sonic mass to the fundamental masses of stars. 

Important questions arise: How sensitive is the high-mass slope of the IMF ($\alpha_{\rm high}$) to the environment?  In the end, how strongly does the IMF regulate itself in the process of thermal feedback?  \newtext{What limitations arise from our physical and dynamical assumptions?  And, how well do our predictions match trends inferred within well-studied globular clusters?   }

\subsection{High-mass slope and primordial segregation} \label{SS:alpha_high} 
We consider massive stars to be the dominant fragments in sub-regions that collapse from the turbulent background \cite[e.g.,][]{2004MNRAS.349..735B}.    \newtext{In the excursion-set theory of \cite{2012MNRAS.423.2016H,2012MNRAS.423.2037H}, the mass function of collapsing regions is determined by the statistics first barrier crossings, with high-mass slopes like those of giant molecular clouds ($\alpha_{\rm coll} \simeq 1.7$: \citealt{2012MNRAS.423.2016H} eq.\ 29) while the upper stellar mass function acquires the steeper distribution of second barrier crossings (typically $\alpha_{\rm high}\simeq 2.35$: \citealt{2012MNRAS.423.2037H} eq.\ 31).  The role of sub-fragmentation in steepening $\alpha_{\rm high}$ was also stressed by  \citet{2011ApJ...739L..46O}.  } 

\newtext{As discussed in \S\,\ref{SS:Mupper_stab}, the tendency of a collapsing region to fragment will depend on its thermal state, as represented by  the accretion stability parameter $\xifil$.  Equation (\ref{eq:xiFil_vs_MonMs}) shows that an increase of $ m_s$  will tend to suppress fragmentation in the regions that collapse to form massive stars, thereby reducing $\alpha_{\rm high}$ to some degree.   If the dependence on $m_s$ is sufficiently strong, it may be possible for this effect to increase the specific luminosity $(L/M)_*$ enough to achieve $\Gamma>1$.  

To explore this possibility, we entertain a relatively strong realization of this effect, such that $\alpha_{\rm high}=2.35$ when $m_{s,\rm env}\simeq 0.2\,M_\odot$ (as in local star formation), flattening to $\alpha_{\rm high}=1.75$ when $m_{s,\rm env} \rightarrow 3.1\,M_\odot$ (the saturated value of eq.\ [\ref{eq:Msonic_Env}]).  This suggests}
\begin{equation}\label{eq:alpha_high_guess}
 \alpha_{\rm high} \simeq 2.0 - 0.5 \log_{10}\left({m_{s,{\rm env}}}/{M_\odot}\right).
 \end{equation} 

 \newtext{While this specific form is purely hypothetical, the existence of a positive correlation between $m_s$ and $\alpha_{\rm high}$ is supported by the ability of stellar radiation to inhibit fragmentation in the simulations of \cite{2011ApJ...740...74K}, \cite{2011ApJ...742L...9C}, and \cite{2013ApJ...766...97M}, for instance.  Such a correlation also promotes a tendency for higher-column regions to host more massive stars, and for the primordial segregation of massive stars toward the high-column regions within an optically thick clump, as we discuss below.  If the correlation is sufficiently strong to generate $(L/M)_{*3}\gtrsim 2$, as equation (\ref{eq:alpha_high_guess}) is, then it can self-consistently lead to $\Gamma\gtrsim1$ within optically thick regions.  This allows for radiation pressure to dynamically regulate star formation \citep[e.g.,][]{2018MNRAS.481.4895C}, as we demonstrate  using a toy model in \S\,\ref{SS:toymodel}.  } 
 
\newtext{It is straightforward to see why a positive correlation implies that massive stars form preferentially in the thickest parts of optically thick regions.} 
In the optically thick regime,  $m_{s,\rm env}$ increases nearly linearly with  $\deltagr\Sigma (L/M)_*$ so long as $T<150$\,K (eq.\ [\ref{eq:ms_env}]).   An increase in  $m_{s,\rm env}$ will  raise $(L/M)_*$ to some degree, especially if it causes $\alpha_{\rm high}$ to increase as well. 
 This implies that thick regions higher column densities should display flatter mass functions -- and the trend should hold both between regions and within an individual region. In other words, it implies a flattening of the IMF in star clusters of higher initial $(Z/Z_\odot)\Sigma_*$ (among those that are optically thick to begin with), as well a flattening of the IMF toward the high-column portion of a single region. The latter effect amounts to a {primordial segregation} of massive stars within the region.    Note: these phenomena are related to, but somewhat distinct from, the column density threshold proposed by \citealt{2008Natur.451.1082K}.  It is likely that their threshold plays a role in preventing massive star formation in our optically thin regime.  
 
 For regions that warm themselves above 150\,K, the dust opacity becomes roughly constant, weakening somewhat the sensitivity of  $m_{s,{\rm env}}$ and $\alpha_{\rm high}$ to the environmental parameters.

\begin{figure*}[ht!] 
\begin{centering} 
\includegraphics[scale=0.74]{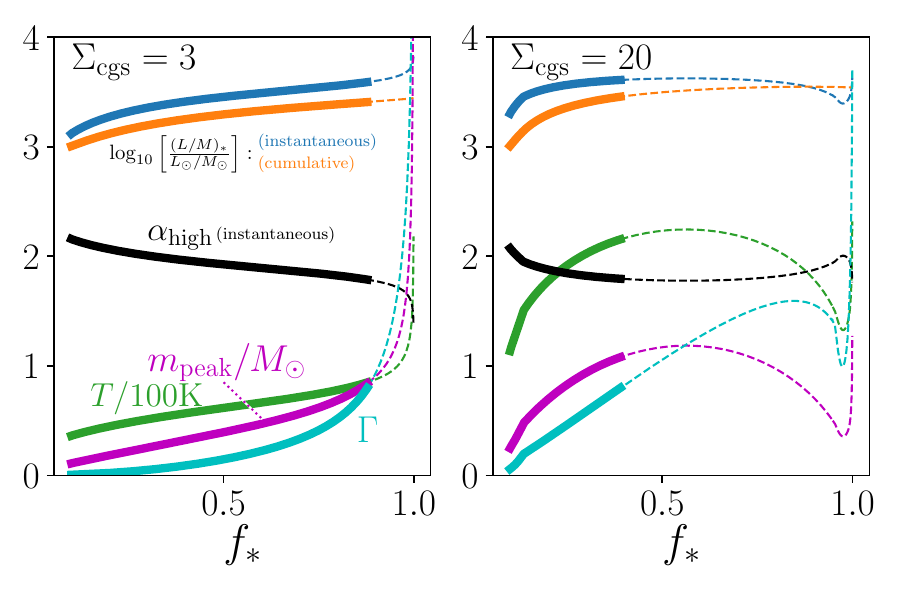}
\caption{A toy model for star cluster formation, in which $\Sigma$ is held fixed at either 3 (left) or 20 (right), and the stellar population evolves under the influence of its own radiation \newtext{while the upper-mass slope evolves according to the hypothetical relation (\ref{eq:alpha_high_guess}). }   Curves are plotted as dashed lines once the Eddington factor $\Gamma$ exceeds 0.8, as this is likely to imply disruption by radiation pressure gradients.  In the right panel, the kink in the lines arises from the change in the opacity law at $T=150$\,K.    See \S\,\ref{SS:toymodel} for additional details.  }
\label{Fig:ToyModel}
\end{centering}
\end{figure*}
 
\subsection{A toy model}\label{SS:toymodel} 
A toy model of star cluster formation helps to illustrate some of these interactions, \newtext{and demonstrate that a relation like equation (\ref{eq:alpha_high_guess}) can lead to dynamically significant radiation pressure.}  We stress that the toy model presented here is not meant to represent a realistic evolution.   We grow a stellar system by holding $\Sigma$ artificially fixed, building up the luminosity and mass of a stellar population by evolving from low to high $f_*$.  Ingredients for this include: $(i)$ a piecewise power law model for the [Planck, Rosseland] opacities, $\kappa_{\rm [Pl,R]} = [6.1,3.0]\deltagr \min(T/100, 1.5)^{[1.93,1.58]}$;  $(ii)$ a estimate of $\tau(T)$, for which we use $(\tau_{\rm Pl} + 3\tau_{\rm R}^4)/(1 + 3\tau_{\rm R}^3)$, where $\tau_{\rm [Pl,R]}$ are the self-consistent [Planck, Rosseland] optical depths given $T$, $\deltagr f_g\Sigma$, and $f_* (L/M)_*$;  $(iii)$ a self-consistent value of $T$ given $\deltagr f_g \Sigma$ and $f_* (L/M)_*$, which involves solving equation (\ref{eq:Tapprox})  with $f(\tau) = [\tau_{\rm Pl}^{-2} + (\tau_{\rm R}/\phi_\tau)^2 ]^{1/2}$; $(iv)$ an estimate of $m_{\rm opac}$, as the maximum of the options in equation (\ref{eq:Mmin}); $(v)$ an evaluation of $m_{s,\rm env}$ using equation (\ref{eq:ms_general}).; $(vi)$ an evaluation of $X$ and $\varepsilon$ from equations (\ref{eq:efficX_senv}) and (\ref{eq:epsilon}); and $(vii)$ an estimate of $\alpha_{\rm high}$, for which we use equation (\ref{eq:alpha_high_guess}).   

Given these, we construct at every point in the evolution an instantaneous IMF that has a log-normal section, peaked at $m_{\rm peak} = \phi_{\rm mult} \varepsilon m_{s,\rm env}$ (with $ \phi_{\rm mult} $ fixed at 0.75 for simplicity), has a width of $0.5$ dex, and is bounded below by $m_{\rm min} = m_{\rm opac}$.  This is joined  to a high-mass power law of logarithmic slope $-\alpha_{\rm high}$, in such a way that the IMF and its slope are both continuous. This extends to an upper limit $m_{\rm max}$, held fixed at $120\,M_\odot$ for simplicity.  For this IMF the value of $(L/M)_*$  is estimated using the zero-age main sequence luminosity from \cite{1996MNRAS.281..257T}, which determines $dL_*/dM_*$ as stellar mass is accumulated.  This is a simple integration, because we ignore the possibility that there might be time for stellar evolution to change the luminosity of the extant stars.

Figure \ref{Fig:ToyModel} provides two examples of this toy model, for $\SigmaCGS=3$ and $\SigmaCGS=20$, in which $\alphavir=\deltagr=1$.  In both cases we start with $f_*=0.1$ and $(L/M)_{*,3}=1$ and evolve to higher $f_*$.  In both models, the thermal feedback from newborn stars is sufficient to raise $m_s$ and flatten $\alpha_{\rm high}$ to the extent that the Eddington factor becomes significant ($\Gamma>0.8$, say).  As expected, this transition happens earlier in the higher-column example (at $f_*=0.64$ for $\SigmaCGS=20$, compared to $f_*=0.88$ for $\SigmaCGS=3$).  

 \newtext{Importantly, we note that the appearance of a super-Eddington state in this model is the direct consequence of the flattening of $\alpha_{\rm high}$ we assumed by adopting equation (\ref{eq:alpha_high_guess}).  If instead we  fix $\alpha_{\rm high}=2.35$, we find that  $\Gamma>0.8$ is not achieved until the gas is almost entirely consumed.}

\subsection{Regime of validity} \label{SS:validity} 

\newtext{
A few of the assumptions in this paper limit its regime of validity.  First, dust  must be sufficiently abundant that the collisions between gas and grains will tightly couple their temperatures within sonic cores.   
 Examining  \citet{2000ApJ...534..809O}'s figure 1, we estimate that the dust and gas temperatures agree within 15\% when
 $n_H>10^{4.5} (Z/Z_\odot)^{-1.33}$\,cm$^{-3}$.   Noting that the thermal pressure within a sonic core roughly matches the hydrostatic pressure $\sim 0.7 f_g G\Sigma^2$ of the environment, and that the density threshold is always reached within the optically thin branch of equation (\ref{eq:T}), we find that dust and gas are self-consistently coupled whenever \begin{equation} 
 \frac{Z}{Z_\odot} \gtrsim 10^{-2.4} \left(\frac{\SigmaCGS}{3}\right)^{-1.32}. 
 \end{equation}  
This would include all of the globular clusters in the \cite{1996AJ....112.1487H} catalog, assuming they formed in a compact state with $\SigmaCGS\gtrsim3$. 
 
 Second, because we have ignored the cosmic background radiation in determining the dust temperature, our results are only valid up to the redshift at which the background temperature $2.725(1+z)$\,K approaches the temperature in equation (\ref{eq:T}).  This criterion is most relevant to optically thin environments with high dust content, which are relatively cooler, for which it amounts to \begin{equation} z<17.2 \left[\frac{ (L/M)_{*3} (f_g^{-1}-1) } {\deltagr}\right]^{0.18} - 1. \end{equation}    
  
 Third,  our dust temperature estimate should be adjusted  the clumpy nature of the dust distribution and from incomplete thermalization of the radiation field.  \cite{2018MNRAS.481.4895C} use the characteristics of fully developed supersonic turbulence  to argue that the impact of inhomogeneities on the transfer of radiation will be relatively minor (their Appendix C), but radiation hydrodynamics solutions can help to calibrate this.  Likewise, Monte Carlo transfer solutions will be useful to quantify the impact of a non-thermal radiation field.  
 
 Finally, although we have not adopted a specific cluster formation scenario, we have assumed it is rapid enough that no stars leave the main sequence.   Also,  when estimating $\tau$ we assumed significant gas fraction; this is more consistent with the monolithic collapse than extended accretion.  The formation of enriched stars from cooled stellar ejecta  \citep{2017ApJ...835...60W} or ejecta mixed with ambient material \citep{2017MNRAS.470..977L} involves extended accretion, so this assumption may be more appropriate for the first generation than the second one. 
 } 

\subsection{Application to well-studied globular clusters }

\newtext{  In our theory, the low-mass IMF should depend primarily on the value of $m_s$, with secondary modifications from binary fragmentation and outflow ejection.  It is interesting to compare this  to the apparent IMF variations within globular clusters, bearing in mind that  living stars within them  have $m_*\lesssim 0.8 M_\odot$, and that stellar remnants contribute to the total mass while they linger \citep[e.g.,][]{2013MNRAS.432.2779B}. Qualitatively, the relatively large values of $m_s$ implied by equations (\ref{eq:ms_env}) and (\ref{eq:ms_self}) are consistent with the inferences of bottom-light IMFs mentioned in \S\,\ref{S:Intro}.   Quantitatively, our finding that $m_{s,\rm env}$ is approximately $\propto (\deltagr^{-1/3}\Sigma^{-1}, \deltagr \Sigma)$ in optically (thin, thick) clouds is only compatible with \cite{2011AJ....142....8S}'s determination that clusters in M31 are more bottom light at low $Z$ if the low-mass stars in question form in an optically thin region -- perhaps in their outskirts.  (Note also: \cite{2016ApJ...826...89Z} conclude that $ Z^{-1/7}\rho_*$  correlates with a bottom-light IMF; however \cite{2011AJ....142....8S} attribute the variation with $\rho_*$ to dynamical evolution.)   These points warrant further study.}

\section{Conclusions} \label{S:Conclusions} 
We  draw four major conclusions about  intense starbursts like those that give rise to massive globular clusters: 
\begin{enumerate}
\item Important quantities, such as the opacity-limited minimum mass, the sonic mass, the efficiency of individual star formation, the tendency for binary fragmentation, and the high-mass slope of the IMF, can all be related to the radiation environment, which in turn is determined by the transport of stellar radiation through the dusty star-forming environment.  Each of these quantities can be approximated analytically in the optically thin or thick cases, although some of the forms proposed here ($\alpha_{\rm high}$, in particular) are quite speculative. 
\item When expressed in terms of the Eddington factor $\Gamma$, the sonic mass $m_s$ within an optically thick starburst region is related to the fundamental mass scale composed of the Planck mass and the mean molecular weight.  (This was already true of the minimum stellar mass: \citealt{1976MNRAS.176..483R}.)  Because this is similar to the fundamental mass scales for stellar evolution, such as Chandrasekhar's limit, it implies that the ultimate production of stellar remnants  is not entirely arbitrary.  \newtext{At a basic level, this  arises because radiation and gas pressures become comparable within sonic cores.} 
\item Among regions that are optically thick to their own thermal radiation, a plausible line of argument suggests that the IMF should flatten as the sonic mass increases, and that this  accompanies an increase of the quantity $(Z/Z_\odot)\Sigma$.   This phenomenon is expected to operate among star forming environments and also within them, where it produces a primordial segregation of massive stars in the highest-column substructures.  
\item  The Eddington factor $\Gamma$ can become appreciable in the midst of star cluster formation, thanks to the luminosity-to-mass ratio of young stellar populations and the character of dust opacities.  In particular, a relatively small flattening of the IMF is sufficient to allow radiation pressure to become dynamically significant.   We present a toy model in which this occurs naturally in the process of star cluster formation, \newtext{while also noting that the sensitivity of the IMF slope to thermal feedback has a controlling influence on the outcome.}
\end{enumerate}

This work complements numerical studies of the IMF that explore the importance of thermal feedback, such as those of  \cite{2021MNRAS.508.4175C}, \cite{2022MNRAS.515.4929G}, \cite{2022MNRAS.516.5712T},  and \cite{2024MNRAS.527.7306T}, as well as studies that address the dynamical influence of radiation pressure, such as \cite{kim2018}, \cite{2018MNRAS.478...81C}, \cite{2018MNRAS.481.4895C},  and \cite{2024ApJ...967L..28M}.    An analytical approach is  useful, because of the very high dynamic range required to simulate regions as massive as globular clusters with the mass resolution required to capture fragmentation.

\section*{Acknowledgements}

Very helpful comments from Chris McKee, Liang Dai, Scott Tremaine, Massimo Pascale, James Gurian,  Kaitlin Kratter, Sally Oey,  Mike Grudi\'c, and the referee are all deeply appreciated. 
  This research has been supported in part by an NSERC Discovery Grant. 

%\end{acknowledgements}

\bibliographystyle{apj}
%\bibliography{StarFormation}
\bibliography{IMFpaper}

\begin{thebibliography}{92}
\expandafter\ifx\csname natexlab\endcsname\relax\def\natexlab#1{#1}\fi

\bibitem[{{Bate}(2009)}]{2009MNRAS.392.1363B}
{Bate}, M.~R. 2009, \mnras, 392, 1363

\bibitem[{{Baumgardt} {et~al.}(2008){Baumgardt}, {De Marchi}, \&
  {Kroupa}}]{2008ApJ...685..247B}
{Baumgardt}, H., {De Marchi}, G., \& {Kroupa}, P. 2008, \apj, 685, 247

\bibitem[{{Baumgardt} {et~al.}(2023){Baumgardt}, {H{\'e}nault-Brunet},
  {Dickson}, \& {Sollima}}]{2023MNRAS.521.3991B}
{Baumgardt}, H., {H{\'e}nault-Brunet}, V., {Dickson}, N., \& {Sollima}, A.
  2023, \mnras, 521, 3991

\bibitem[{{Bertoldi} \& {McKee}(1992)}]{1992ApJ...395..140B}
{Bertoldi}, F. \& {McKee}, C.~F. 1992, \apj, 395, 140

\bibitem[{{Bonnell} {et~al.}(2004){Bonnell}, {Vine}, \&
  {Bate}}]{2004MNRAS.349..735B}
{Bonnell}, I.~A., {Vine}, S.~G., \& {Bate}, M.~R. 2004, \mnras, 349, 735

\bibitem[{{Bonnor}(1956)}]{1956MNRAS.116..351B}
{Bonnor}, W.~B. 1956, \mnras, 116, 351

\bibitem[{{Breen} \& {Heggie}(2013)}]{2013MNRAS.432.2779B}
{Breen}, P.~G. \& {Heggie}, D.~C. 2013, \mnras, 432, 2779

\bibitem[{{Burkert} \& {Bodenheimer}(2000)}]{2000ApJ...543..822B}
{Burkert}, A. \& {Bodenheimer}, P. 2000, \apj, 543, 822

\bibitem[{{Cameron} {et~al.}(2023){Cameron}, {Katz}, {Witten}, {Saxena},
  {Laporte}, \& {Bunker}}]{2023arXiv231102051C}
{Cameron}, A.~J., {Katz}, H., {Witten}, C., {Saxena}, A., {Laporte}, N., \&
  {Bunker}, A.~J. 2023, arXiv e-prints, arXiv:2311.02051

\bibitem[{{Chabrier}(2003)}]{2003PASP..115..763C}
{Chabrier}, G. 2003, \pasp, 115, 763

\bibitem[{{Chabrier}(2005)}]{2005ASSL..327...41C}
{Chabrier}, G. 2005, in Astrophysics and Space Science Library, Vol. 327, The
  Initial Mass Function 50 Years Later, ed. E.~{Corbelli}, F.~{Palla}, \&
  H.~{Zinnecker}, 41

\bibitem[{{Chabrier} \& {Dumond}(2024)}]{2024ApJ...966...48C}
{Chabrier}, G. \& {Dumond}, P. 2024, \apj, 966, 48

\bibitem[{{Chakrabarti} \& {McKee}(2005)}]{2005ApJ...631..792C}
{Chakrabarti}, S. \& {McKee}, C.~F. 2005, \apj, 631, 792

\bibitem[{{Chandrasekhar}(1931)}]{1931ApJ....74...81C}
{Chandrasekhar}, S. 1931, \apj, 74, 81

\bibitem[{{Chon} {et~al.}(2021){Chon}, {Omukai}, \&
  {Schneider}}]{2021MNRAS.508.4175C}
{Chon}, S., {Omukai}, K., \& {Schneider}, R. 2021, \mnras, 508, 4175

\bibitem[{{Commer{\c{c}}on} {et~al.}(2011){Commer{\c{c}}on}, {Hennebelle}, \&
  {Henning}}]{2011ApJ...742L...9C}
{Commer{\c{c}}on}, B., {Hennebelle}, P., \& {Henning}, T. 2011, \apjl, 742, L9

\bibitem[{{Crocker} {et~al.}(2018{\natexlab{a}}){Crocker}, {Krumholz},
  {Thompson}, {Baumgardt}, \& {Mackey}}]{2018MNRAS.481.4895C}
{Crocker}, R.~M., {Krumholz}, M.~R., {Thompson}, T.~A., {Baumgardt}, H., \&
  {Mackey}, D. 2018{\natexlab{a}}, \mnras, 481, 4895

\bibitem[{{Crocker} {et~al.}(2018{\natexlab{b}}){Crocker}, {Krumholz},
  {Thompson}, \& {Clutterbuck}}]{2018MNRAS.478...81C}
{Crocker}, R.~M., {Krumholz}, M.~R., {Thompson}, T.~A., \& {Clutterbuck}, J.
  2018{\natexlab{b}}, \mnras, 478, 81

\bibitem[{{Dib}(2023)}]{2023ApJ...959...88D}
{Dib}, S. 2023, \apj, 959, 88

\bibitem[{{Ebert}(1955)}]{1955ZA.....37..217E}
{Ebert}, R. 1955, \zap, 37, 217

\bibitem[{{Eddington}(1920)}]{1920Obs....43..341E}
{Eddington}, A.~S. 1920, The Observatory, 43, 341

\bibitem[{{El-Badry} \& {Rix}(2019)}]{2019MNRAS.482L.139E}
{El-Badry}, K. \& {Rix}, H.-W. 2019, \mnras, 482, L139

\bibitem[{{Elmegreen} {et~al.}(2008){Elmegreen}, {Klessen}, \&
  {Wilson}}]{2008ApJ...681..365E}
{Elmegreen}, B.~G., {Klessen}, R.~S., \& {Wilson}, C.~D. 2008, \apj, 681, 365

\bibitem[{{Fall} {et~al.}(2010){Fall}, {Krumholz}, \&
  {Matzner}}]{2010ApJ...710L.142F}
{Fall}, S.~M., {Krumholz}, M.~R., \& {Matzner}, C.~D. 2010, \apjl, 710, L142

\bibitem[{{Fiege} \& {Pudritz}(2000)}]{2000MNRAS.311..105F}
{Fiege}, J.~D. \& {Pudritz}, R.~E. 2000, \mnras, 311, 105

\bibitem[{{Fujimoto} {et~al.}(2023){Fujimoto}, {Wang}, {Weaver}, {Kokorev},
  {Atek}, {Bezanson}, {Labbe}, {Brammer}, {Greene}, {Chemerynska}, {Dayal}, {de
  Graaff}, {Furtak}, {Oesch}, {Setton}, {Price}, {Miller}, {Williams},
  {Whitaker}, {Zitrin}, {Cutler}, {Leja}, {Pan}, {Coe}, {van Dokkum},
  {Feldmann}, {Fudamoto}, {Goulding}, {Khullar}, {Marchesini}, {Maseda},
  {Nanayakkara}, {Nelson}, {Smit}, {Stefanon}, \&
  {Weibel}}]{2023arXiv230811609F}
{Fujimoto}, S., {Wang}, B., {Weaver}, J., {Kokorev}, V., {Atek}, H.,
  {Bezanson}, R., {Labbe}, I., {Brammer}, G., {Greene}, J.~E., {Chemerynska},
  I., {Dayal}, P., {de Graaff}, A., {Furtak}, L.~J., {Oesch}, P.~A., {Setton},
  D.~J., {Price}, S.~H., {Miller}, T.~B., {Williams}, C.~C., {Whitaker}, K.~E.,
  {Zitrin}, A., {Cutler}, S.~E., {Leja}, J., {Pan}, R., {Coe}, D., {van
  Dokkum}, P., {Feldmann}, R., {Fudamoto}, Y., {Goulding}, A.~D., {Khullar},
  G., {Marchesini}, D., {Maseda}, M., {Nanayakkara}, T., {Nelson}, E.~J.,
  {Smit}, R., {Stefanon}, M., \& {Weibel}, A. 2023, arXiv e-prints,
  arXiv:2308.11609

\bibitem[{{Gammie}(2001)}]{2001ApJ...553..174G}
{Gammie}, C.~F. 2001, \apj, 553, 174

\bibitem[{{Goldsmith}(2001)}]{2001ApJ...557..736G}
{Goldsmith}, P.~F. 2001, \apj, 557, 736

\bibitem[{{Gratton} {et~al.}(2012){Gratton}, {Carretta}, \&
  {Bragaglia}}]{2012A&ARv..20...50G}
{Gratton}, R.~G., {Carretta}, E., \& {Bragaglia}, A. 2012, \aapr, 20, 50

\bibitem[{{Guszejnov} {et~al.}(2022){Guszejnov}, {Grudi{\'c}}, {Offner},
  {Faucher-Gigu{\`e}re}, {Hopkins}, \& {Rosen}}]{2022MNRAS.515.4929G}
{Guszejnov}, D., {Grudi{\'c}}, M.~Y., {Offner}, S. S.~R.,
  {Faucher-Gigu{\`e}re}, C.-A., {Hopkins}, P.~F., \& {Rosen}, A.~L. 2022,
  \mnras, 515, 4929

\bibitem[{Guszejnov \& Hopkins(2015)}]{10.1093/mnras/stv872}
Guszejnov, D. \& Hopkins, P.~F. 2015, MNRAS, 450, 4137

\bibitem[{{Harris}(1996)}]{1996AJ....112.1487H}
{Harris}, W.~E. 1996, \aj, 112, 1487

\bibitem[{{Hennebelle} \& {Chabrier}(2013)}]{2013ApJ...770..150H}
{Hennebelle}, P. \& {Chabrier}, G. 2013, \apj, 770, 150

\bibitem[{{Hennebelle} {et~al.}(2020){Hennebelle}, {Commer{\c{c}}on}, {Lee}, \&
  {Chabrier}}]{2020ApJ...904..194H}
{Hennebelle}, P., {Commer{\c{c}}on}, B., {Lee}, Y.-N., \& {Chabrier}, G. 2020,
  \apj, 904, 194

\bibitem[{{Hopkins}(2012{\natexlab{a}})}]{2012MNRAS.423.2016H}
{Hopkins}, P.~F. 2012{\natexlab{a}}, \mnras, 423, 2016

\bibitem[{{Hopkins}(2012{\natexlab{b}})}]{2012MNRAS.423.2037H}
---. 2012{\natexlab{b}}, \mnras, 423, 2037

\bibitem[{Hopkins {et~al.}(2010)Hopkins, Murray, Quataert, \&
  Thompson}]{hopkins2010maximum}
Hopkins, P.~F., Murray, N., Quataert, E., \& Thompson, T.~A. 2010, Monthly
  Notices of the Royal Astronomical Society: Letters, 401, L19

\bibitem[{{Hosek} {et~al.}(2019){Hosek}, {Lu}, {Anderson}, {Najarro}, {Ghez},
  {Morris}, {Clarkson}, \& {Albers}}]{2019ApJ...870...44H}
{Hosek}, Matthew~W., J., {Lu}, J.~R., {Anderson}, J., {Najarro}, F., {Ghez},
  A.~M., {Morris}, M.~R., {Clarkson}, W.~I., \& {Albers}, S.~M. 2019, \apj,
  870, 44

\bibitem[{{Kenyon} {et~al.}(1990){Kenyon}, {Hartmann}, {Strom}, \&
  {Strom}}]{1990AJ.....99..869K}
{Kenyon}, S.~J., {Hartmann}, L.~W., {Strom}, K.~M., \& {Strom}, S.~E. 1990,
  \aj, 99, 869

\bibitem[{Kim {et~al.}(2018)Kim, Kim, \& Ostriker}]{kim2018}
Kim, J.-G., Kim, W.-T., \& Ostriker, E.~C. 2018, The Astrophysical Journal,
  859, 68

\bibitem[{{Klassen} {et~al.}(2017){Klassen}, {Pudritz}, \&
  {Kirk}}]{2017MNRAS.465.2254K}
{Klassen}, M., {Pudritz}, R.~E., \& {Kirk}, H. 2017, \mnras, 465, 2254

\bibitem[{{Kratter} \& {Matzner}(2006)}]{2006MNRAS.373.1563K}
{Kratter}, K.~M. \& {Matzner}, C.~D. 2006, \mnras, 373, 1563

\bibitem[{{Kratter} {et~al.}(2010){Kratter}, {Matzner}, {Krumholz}, \&
  {Klein}}]{2010ApJ...708.1585K}
{Kratter}, K.~M., {Matzner}, C.~D., {Krumholz}, M.~R., \& {Klein}, R.~I. 2010,
  \apj, 708, 1585

\bibitem[{{Kroupa}(2001)}]{2001MNRAS.322..231K}
{Kroupa}, P. 2001, \mnras, 322, 231

\bibitem[{{Krumholz}(2011)}]{2011ApJ...743..110K}
{Krumholz}, M.~R. 2011, \apj, 743, 110

\bibitem[{{Krumholz} {et~al.}(2011){Krumholz}, {Klein}, \&
  {McKee}}]{2011ApJ...740...74K}
{Krumholz}, M.~R., {Klein}, R.~I., \& {McKee}, C.~F. 2011, \apj, 740, 74

\bibitem[{{Krumholz} \& {McKee}(2008)}]{2008Natur.451.1082K}
{Krumholz}, M.~R. \& {McKee}, C.~F. 2008, \nat, 451, 1082

\bibitem[{Krumholz {et~al.}(2004)Krumholz, McKee, \&
  Klein}]{krumholz2004protostellar}
Krumholz, M.~R., McKee, C.~F., \& Klein, R.~I. 2004, The Astrophysical Journal,
  618, L33

\bibitem[{{Lee} \& {Hennebelle}(2018)}]{2018A&A...611A..89L}
{Lee}, Y.-N. \& {Hennebelle}, P. 2018, \aap, 611, A89

\bibitem[{{Lochhaas} \& {Thompson}(2017)}]{2017MNRAS.470..977L}
{Lochhaas}, C. \& {Thompson}, T.~A. 2017, \mnras, 470, 977

\bibitem[{{Low} \& {Lynden-Bell}(1976)}]{1976MNRAS.176..367L}
{Low}, C. \& {Lynden-Bell}, D. 1976, \mnras, 176, 367

\bibitem[{{Lu} {et~al.}(2013){Lu}, {Do}, {Ghez}, {Morris}, {Yelda}, \&
  {Matthews}}]{2013ApJ...764..155L}
{Lu}, J.~R., {Do}, T., {Ghez}, A.~M., {Morris}, M.~R., {Yelda}, S., \&
  {Matthews}, K. 2013, \apj, 764, 155

\bibitem[{{Marks} \& {Kroupa}(2010)}]{2010MNRAS.406.2000M}
{Marks}, M. \& {Kroupa}, P. 2010, \mnras, 406, 2000

\bibitem[{{Marques-Chaves} {et~al.}(2024){Marques-Chaves}, {Schaerer},
  {Kuruvanthodi}, {Korber}, {Prantzos}, {Charbonnel}, {Weibel}, {Izotov},
  {Messa}, {Brammer}, {Dessauges-Zavadsky}, \& {Oesch}}]{2024A&A...681A..30M}
{Marques-Chaves}, R., {Schaerer}, D., {Kuruvanthodi}, A., {Korber}, D.,
  {Prantzos}, N., {Charbonnel}, C., {Weibel}, A., {Izotov}, Y.~I., {Messa}, M.,
  {Brammer}, G., {Dessauges-Zavadsky}, M., \& {Oesch}, P. 2024, \aap, 681, A30

\bibitem[{{Masunaga} \& {Inutsuka}(1999)}]{1999ApJ...510..822M}
{Masunaga}, H. \& {Inutsuka}, S.-i. 1999, \apj, 510, 822

\bibitem[{{Matzner} \& {Jumper}(2015)}]{2015ApJ...815...68M}
{Matzner}, C.~D. \& {Jumper}, P.~H. 2015, \apj, 815, 68

\bibitem[{{Matzner} \& {Levin}(2005)}]{2005ApJ...628..817M}
{Matzner}, C.~D. \& {Levin}, Y. 2005, \apj, 628, 817

\bibitem[{{Matzner} \& {McKee}(2000)}]{2000ApJ...545..364M}
{Matzner}, C.~D. \& {McKee}, C.~F. 2000, \apj, 545, 364

\bibitem[{{McKee} \& {Zweibel}(1995)}]{1995ApJ...440..686M}
{McKee}, C.~F. \& {Zweibel}, E.~G. 1995, \apj, 440, 686

\bibitem[{{Menon} {et~al.}(2024){Menon}, {Lancaster}, {Burkhart}, {Somerville},
  {Dekel}, \& {Krumholz}}]{2024ApJ...967L..28M}
{Menon}, S.~H., {Lancaster}, L., {Burkhart}, B., {Somerville}, R.~S., {Dekel},
  A., \& {Krumholz}, M.~R. 2024, \apjl, 967, L28

\bibitem[{{Moe} {et~al.}(2019){Moe}, {Kratter}, \&
  {Badenes}}]{2019ApJ...875...61M}
{Moe}, M., {Kratter}, K.~M., \& {Badenes}, C. 2019, \apj, 875, 61

\bibitem[{{Myers} {et~al.}(2013){Myers}, {McKee}, {Cunningham}, {Klein}, \&
  {Krumholz}}]{2013ApJ...766...97M}
{Myers}, A.~T., {McKee}, C.~F., {Cunningham}, A.~J., {Klein}, R.~I., \&
  {Krumholz}, M.~R. 2013, \apj, 766, 97

\bibitem[{{Navarro-Carrera} {et~al.}(2024){Navarro-Carrera}, {Caputi}, {Iani},
  {Rinaldi}, {Kokorev}, \& {Kerutt}}]{2024arXiv240714201N}
{Navarro-Carrera}, R., {Caputi}, K.~I., {Iani}, E., {Rinaldi}, P., {Kokorev},
  V., \& {Kerutt}, J. 2024, arXiv e-prints, arXiv:2407.14201

\bibitem[{{Nayakshin} \& {Sunyaev}(2005)}]{2005MNRAS.364L..23N}
{Nayakshin}, S. \& {Sunyaev}, R. 2005, \mnras, 364, L23

\bibitem[{{Oey}(2011)}]{2011ApJ...739L..46O}
{Oey}, M.~S. 2011, \apjl, 739, L46

\bibitem[{{Offner} {et~al.}(2009){Offner}, {Klein}, {McKee}, \&
  {Krumholz}}]{2009ApJ...703..131O}
{Offner}, S. S.~R., {Klein}, R.~I., {McKee}, C.~F., \& {Krumholz}, M.~R. 2009,
  \apj, 703, 131

\bibitem[{{Offner} {et~al.}(2010){Offner}, {Kratter}, {Matzner}, {Krumholz}, \&
  {Klein}}]{2010ApJ...725.1485O}
{Offner}, S. S.~R., {Kratter}, K.~M., {Matzner}, C.~D., {Krumholz}, M.~R., \&
  {Klein}, R.~I. 2010, \apj, 725, 1485

\bibitem[{{Offner} \& {McKee}(2011)}]{2011ApJ...736...53O}
{Offner}, S. S.~R. \& {McKee}, C.~F. 2011, \apj, 736, 53

\bibitem[{{Offner} {et~al.}(2023){Offner}, {Moe}, {Kratter}, {Sadavoy},
  {Jensen}, \& {Tobin}}]{2023ASPC..534..275O}
{Offner}, S.~S.~R., {Moe}, M., {Kratter}, K.~M., {Sadavoy}, S.~I., {Jensen},
  E.~L.~N., \& {Tobin}, J.~J. 2023, in Astronomical Society of the Pacific
  Conference Series, Vol. 534, Protostars and Planets VII, ed. S.~{Inutsuka},
  Y.~{Aikawa}, T.~{Muto}, K.~{Tomida}, \& M.~{Tamura}, 275

\bibitem[{{Omukai}(2000)}]{2000ApJ...534..809O}
{Omukai}, K. 2000, \apj, 534, 809

\bibitem[{{Ostriker}(1964)}]{1964ApJ...140.1056O}
{Ostriker}, J. 1964, \apj, 140, 1056

\bibitem[{{Palla} \& {Stahler}(1990)}]{1990ApJ...360L..47P}
{Palla}, F. \& {Stahler}, S.~W. 1990, \apjl, 360, L47

\bibitem[{{Pascale} \& {Dai}(2024)}]{2024arXiv240410755P}
{Pascale}, M. \& {Dai}, L. 2024, arXiv e-prints, arXiv:2404.10755

\bibitem[{{Pascale} {et~al.}(2023){Pascale}, {Dai}, {McKee}, \&
  {Tsang}}]{2023ApJ...957...77P}
{Pascale}, M., {Dai}, L., {McKee}, C.~F., \& {Tsang}, B. T.~H. 2023, \apj, 957,
  77

\bibitem[{{Rees}(1976)}]{1976MNRAS.176..483R}
{Rees}, M.~J. 1976, \mnras, 176, 483

\bibitem[{{Robitaille}(2017)}]{2017A&A...600A..11R}
{Robitaille}, T.~P. 2017, \aap, 600, A11

\bibitem[{{Schaerer} {et~al.}(2024){Schaerer}, {Marques-Chaves}, {Xiao}, \&
  {Korber}}]{2024arXiv240608408S}
{Schaerer}, D., {Marques-Chaves}, R., {Xiao}, M., \& {Korber}, D. 2024, arXiv
  e-prints, arXiv:2406.08408

\bibitem[{{Semenov} {et~al.}(2003){Semenov}, {Henning}, {Helling}, {Ilgner}, \&
  {Sedlmayr}}]{2003A&A...410..611S}
{Semenov}, D., {Henning}, T., {Helling}, C., {Ilgner}, M., \& {Sedlmayr}, E.
  2003, \aap, 410, 611

\bibitem[{{Smith} {et~al.}(2023){Smith}, {Oey}, {Hernandez}, {Ryon},
  {Leitherer}, {Charlot}, {Bruzual}, {Calzetti}, {Chu}, {Hayes}, {James},
  {Jaskot}, \& {{\"O}stlin}}]{2023ApJ...958..194S}
{Smith}, L.~J., {Oey}, M.~S., {Hernandez}, S., {Ryon}, J., {Leitherer}, C.,
  {Charlot}, S., {Bruzual}, G., {Calzetti}, D., {Chu}, Y.-H., {Hayes}, M.~J.,
  {James}, B.~L., {Jaskot}, A.~E., \& {{\"O}stlin}, G. 2023, \apj, 958, 194

\bibitem[{{Stamatellos} \& {Whitworth}(2009)}]{2009MNRAS.400.1563S}
{Stamatellos}, D. \& {Whitworth}, A.~P. 2009, \mnras, 400, 1563

\bibitem[{{Strader} {et~al.}(2011){Strader}, {Caldwell}, \&
  {Seth}}]{2011AJ....142....8S}
{Strader}, J., {Caldwell}, N., \& {Seth}, A.~C. 2011, \aj, 142, 8

\bibitem[{{Tanvir} \& {Krumholz}(2024)}]{2024MNRAS.527.7306T}
{Tanvir}, T.~S. \& {Krumholz}, M.~R. 2024, \mnras, 527, 7306

\bibitem[{{Tanvir} {et~al.}(2022){Tanvir}, {Krumholz}, \&
  {Federrath}}]{2022MNRAS.516.5712T}
{Tanvir}, T.~S., {Krumholz}, M.~R., \& {Federrath}, C. 2022, \mnras, 516, 5712

\bibitem[{{Toomre}(1964)}]{1964ApJ...139.1217T}
{Toomre}, A. 1964, \apj, 139, 1217

\bibitem[{{Topping} {et~al.}(2024){Topping}, {Stark}, {Senchyna}, {Chen},
  {Zitrin}, {Endsley}, {Charlot}, {Furtak}, {Maseda}, {Plat}, {Smit},
  {Mainali}, {Chevallard}, {Molyneux}, \& {Rigby}}]{2024arXiv240719009T}
{Topping}, M.~W., {Stark}, D.~P., {Senchyna}, P., {Chen}, Z., {Zitrin}, A.,
  {Endsley}, R., {Charlot}, S., {Furtak}, L.~J., {Maseda}, M.~V., {Plat}, A.,
  {Smit}, R., {Mainali}, R., {Chevallard}, J., {Molyneux}, S., \& {Rigby},
  J.~R. 2024, arXiv e-prints, arXiv:2407.19009

\bibitem[{{Tout} {et~al.}(1996){Tout}, {Pols}, {Eggleton}, \&
  {Han}}]{1996MNRAS.281..257T}
{Tout}, C.~A., {Pols}, O.~R., {Eggleton}, P.~P., \& {Han}, Z. 1996, \mnras,
  281, 257

\bibitem[{{Upadhyaya} {et~al.}(2024){Upadhyaya}, {Marques-Chaves}, {Schaerer},
  {Martins}, {P{\'e}rez-Fournon}, {Palacios}, \&
  {Stanway}}]{2024A&A...686A.185U}
{Upadhyaya}, A., {Marques-Chaves}, R., {Schaerer}, D., {Martins}, F.,
  {P{\'e}rez-Fournon}, I., {Palacios}, A., \& {Stanway}, E.~R. 2024, \aap, 686,
  A185

\bibitem[{{Urban} {et~al.}(2010){Urban}, {Martel}, \&
  {Evans}}]{2010ApJ...710.1343U}
{Urban}, A., {Martel}, H., \& {Evans}, Neal~J., I. 2010, \apj, 710, 1343

\bibitem[{{V{\'a}zquez-Semadeni} {et~al.}(2003){V{\'a}zquez-Semadeni},
  {Ballesteros-Paredes}, \& {Klessen}}]{2003ApJ...585L.131V}
{V{\'a}zquez-Semadeni}, E., {Ballesteros-Paredes}, J., \& {Klessen}, R.~S.
  2003, \apjl, 585, L131

\bibitem[{{Vink}(2023)}]{2023A&A...679L...9V}
{Vink}, J.~S. 2023, \aap, 679, L9

\bibitem[{{W{\"u}nsch} {et~al.}(2017){W{\"u}nsch}, {Palou{\v{s}}},
  {Tenorio-Tagle}, \& {Ehlerov{\'a}}}]{2017ApJ...835...60W}
{W{\"u}nsch}, R., {Palou{\v{s}}}, J., {Tenorio-Tagle}, G., \& {Ehlerov{\'a}},
  S. 2017, \apj, 835, 60

\bibitem[{{Zonoozi} {et~al.}(2016){Zonoozi}, {Haghi}, \&
  {Kroupa}}]{2016ApJ...826...89Z}
{Zonoozi}, A.~H., {Haghi}, H., \& {Kroupa}, P. 2016, \apj, 826, 89

\end{thebibliography}

\end{document}